\documentclass[aip,amsmath,amssymb,reprint,]{revtex4-1}

\usepackage[utf8]{inputenc}
\usepackage{epstopdf}
\usepackage{graphicx}% Include figure files
\usepackage{dcolumn}% Align table columns on decimal point
\usepackage{bm}
\usepackage{natbib}
\usepackage[mathlines]{lineno}% Enable numbering of text and display math
\usepackage{color}
\usepackage[english]{babel}
\usepackage{hyperref}
\hypersetup{colorlinks=true,citecolor={blue},linkcolor={blue},urlcolor={blue}}
\usepackage{units}
\begin{document}

\preprint{AIP/123-QED}

\title[Polariton condensation in a planar microcavity with InGaAs quantum wells]{Polariton condensation in a planar microcavity with InGaAs quantum wells}

\author{Pasquale Cilibrizzi}
\affiliation{Department of Physics and Astronomy, University of Southampton, Southampton, SO17
1BJ, United Kingdom}

\author{Alexis Askitopoulos} %\footnote{corresponding author: Alexis.Askitopoulos@soton.ac.uk}}
\email{Alexis.Askitopoulos@soton.ac.uk}
\affiliation{Department of Physics and Astronomy, University of Southampton, Southampton, SO17
1BJ, United Kingdom}

\author{Matteo Silva}
\affiliation{Department of Physics and Astronomy, University of Southampton, Southampton, SO17
1BJ, United Kingdom}
\affiliation{Dipartimento di Fisica, Universita di Trento, I-38050 Povo, Italy }

\author{Edmund Clarke}
\affiliation{Department of Electronic and Electrical Engineering, University of Sheffield, Mappin Street, Sheffield S1 4JD, United Kingdom}

\author{Joanna M. Zajac}
\affiliation{School of Physics and Astronomy, Cardiff University, The Parade, CF24 3AA Cardiff, UK}

\author{Wolfgang Langbein}
\affiliation{School of Physics and Astronomy, Cardiff University, The Parade, CF24 3AA Cardiff, UK}

\author{Pavlos G. Lagoudakis}
\affiliation{Department of Physics and Astronomy, University of Southampton, Southampton, SO17
1BJ, United Kingdom}

\date{\today}
\begin{abstract}
Polariton lattice condensates provide a platform for on chip quantum emulations. Interactions in extended polariton lattices are currently limited by the intrinsic photonic disorder of microcavities. Here, we fabricate a strain compensated planar GaAs/AlAs microcavity with embedded InGaAs quantum wells and report on polariton condensation under non-resonant optical excitation. Evidence of polariton condensation is supported spectroscopically both in reflection and transmission geometry, whilst the observation of a second threshold to photon lasing allows us to conclusively distinguish between the strong- and weak-coupling non-linear regimes.
\end{abstract}


\maketitle

Exciton-polaritons are bosonic light-matter quasi-particles formed from the strong coupling between quantum well (QW) excitons and the photonic cavity mode of a semiconductor microcavity (MC) \cite{bookmc}. Increasing the polariton population, the bosonic build up of particles in the ground state of the dispersion gives rise to an inversion-less amplification of the polariton emission \cite{nonequilibrium_1996}. 
This ground state population build up has been widely confirmed as the solid state analogue of Bose-Einstein condensation (BEC) \cite{kasprzak2006}. Solid state polariton condensates have been used to explore fascinating concepts such as superfluidity and quantum vortices in polariton fluidics \cite{amo_collective,KLagoudakis}. 
The BEC phase transition has been demonstrated in a wide range of materials. Condensation at room temperature under optical excitation has been reported, already from 2007 in GaN MCs \cite{room2007} and more recently in ZnO \cite{li_2013} and organic systems \cite{daskalakis,room2014}.
In III-V materials, polariton condensation and lasing has been reported in MCs with GaAs QWs \cite{gaas2009}. 
The implementation of electrically injected polariton condensates has also been reported in GaAs systems \cite{electricallynature,solidstateeinjection} and recently in a GaN microcavity at room temperature\cite{electrical}. From these materials the class of GaAs microcavities is currently the preferred system for the study of polariton fluidics due to minimal photonic and QW disorder and highly controlled fabrication processes. GaAs-based MCs can be further optimised by engineering strain compensation in the cavity mirrors \cite{suppression_2012}. 

 \begin{figure*}
\includegraphics[scale=0.27]{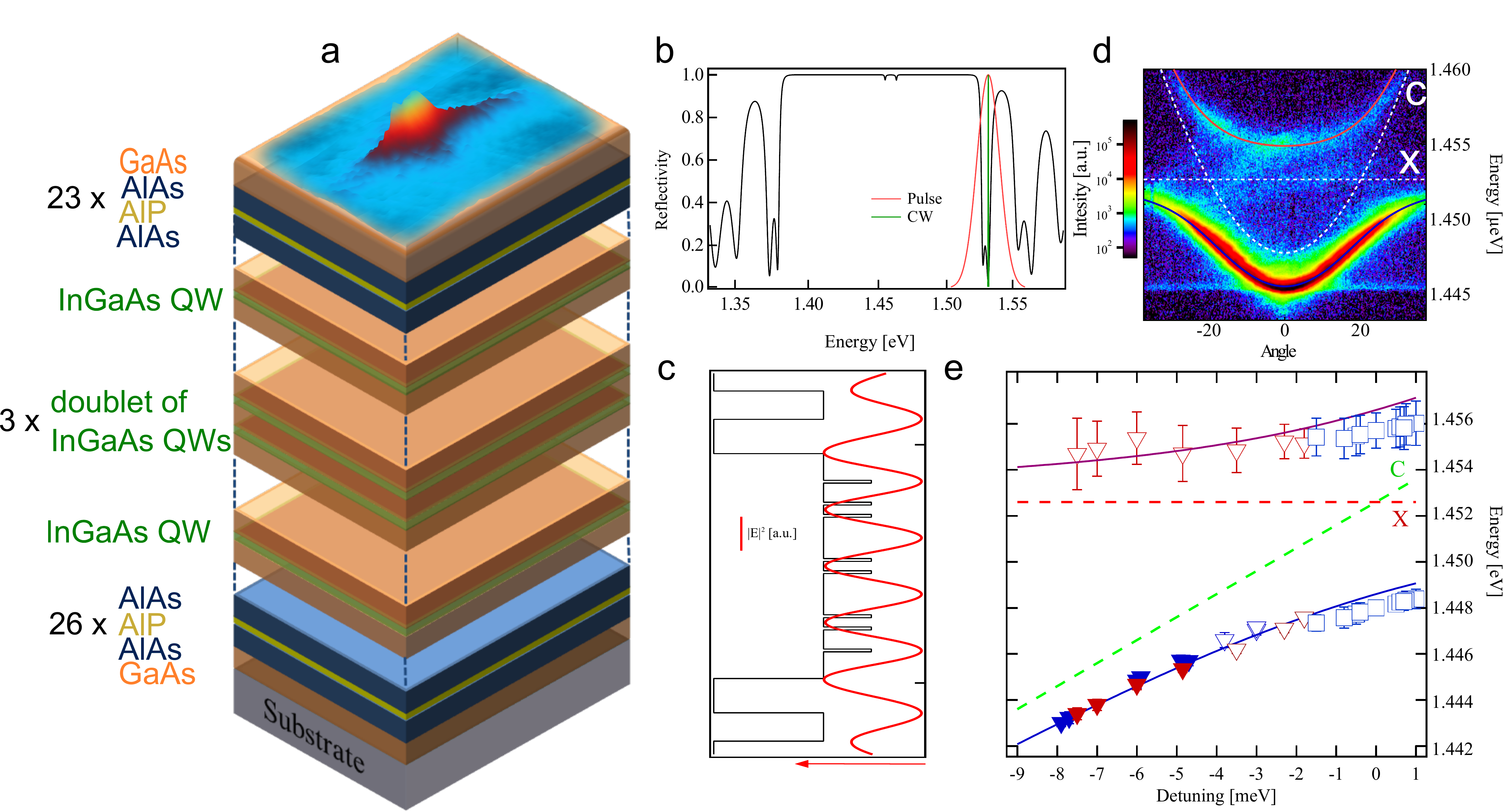}
\centering
\caption{\textbf{a)} Schematic representation of the microcavity structure and condensate emission. \textbf{b)} Calculated reflectivity of the cavity stop band with the transfer matrix method (black line) and corresponding energies of pulsed (red) and CW (green) excitation. \textbf{c)} Schematic of the position of the QWs in the cavity and the corresponding square of the electric field of the cavity mode. \textbf{d)} Polariton dispersion below threshold in a logarithmic false colorscale. The white dashed lines depict the bare exciton and cavity modes and the blue and red solid lines are the calculated upper and lower polariton modes.
\textbf{e)} Upper and lower polariton energy at normal incidence for different detuning conditions (purple, blue lines respectively) and the values recorded from the experiment. The error bars correspond to the FWHM of a gaussian fit to the spectra, while the dashed green (red) line marks the bare cavity (exciton) mode. Square (triangle) symbols mark the reflectivity (transmission) detection geometry, while experiments conducted with pulsed (CW) excitation are shown in red (blue). Filled symbols mark the detuning conditions, where we observe polariton condensation. }
\label{fig1}
\end{figure*}

Here we report on polariton condensation in a planar, strain compensated 2$\lambda$ GaAs/AlAs microcavity with embedded InGaAs QWs under non-resonant optical excitation. Strain compensation is achieved by embedding a $1.3nm$ AlP layer in the center of the AlAs layer of the distributed Bragg reflectors (DBRs). The bottom DBR consists of 26 layers of GaAs and AlAs/AlP/AlAs while the top has 23 pairs as shown in fig.\ref{fig1}a, resulting in very high reflectance ($\textgreater 99.9\%$) in the stop-band region of the spectrum as shown in fig.$\ref{fig1}$b. Three doublets of $6 nm$ $In_{0.08}Ga_{0.92}As$ quantum wells are embedded in the GaAs cavity at the anti-nodes of the field as well as two additional QWs at the first and last $\lambda/4$ of the cavity volume that serve as carrier collection wells and are not strongly coupled as they are at the minimum of the cavity field (fig.$\ref{fig1}c$). The relative large number of QWs in the cavity is expected to maintain the exciton density per QW lower than the Mott density \cite{saturation} and preserve the strong coupling regime to sufficiently high polariton densities to achieve polariton condensation. Strong coupling between the exciton resonance and the cavity mode results in a vacuum rabi-splitting of $\sim8  meV$ (fig.\ref{fig1}d). A wedge in the cavity thickness allows access to a wide range of exciton-cavity detuning conditions (-8 meV to 1 meV). We obtain a lower estimate of the Q-factor of $\sim$ 5500 at -8 meV where the lower polariton branch still has an exciton fraction of 14.6\%.

The experimental setup consists of a low noise cold finger cryostat where the sample is held at $\approx$ 6K and the optical excitation is tuned to the first reflectivity minimum above the cavity stop band as shown in fig.\ref{fig1}b and focused at the top surface of the sample. As the emission energy of the InGaAs QWs is lower than the absorption of the GaAs substrate we can study the photoluminescence of the sample both in reflection and transmission geometry. Extracting the upper and lower polariton energy around $k=0$ for several positions on our sample we construct the energy diagram for different detuning conditions and detection geometries shown in fig.\ref{fig1}e . 

\begin{figure}
\includegraphics[bb=0 0 820 700,scale=0.236]{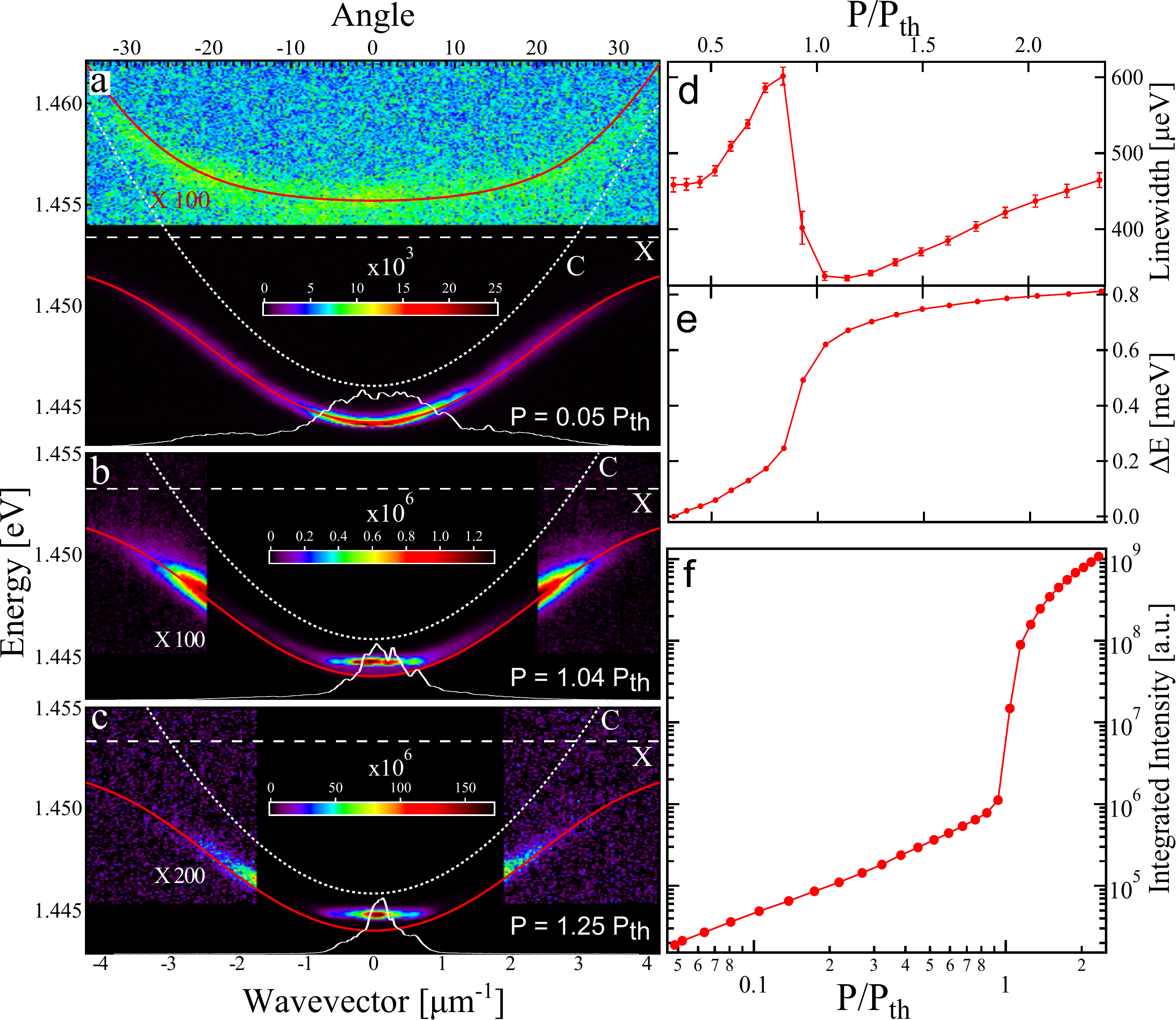}
\centering
\caption{Lower polariton dispersions below \textbf{(a)}, at \textbf{(b)} and above threshold \textbf{(c)}. C and X mark the cavity and exciton mode respectively. In \textbf{(a)} the upper polariton mode is also visible where the intensity is scaled up by 100 times. In \textbf{(b),(c)} we have magnified the intensity of the higher wavevectors by 100 and 200 respectively to clearly show that the excited states follow the polariton dispersion The insets are the corresponding $K-space$ profiles. Linewidth  narrowing \textbf{d)}, energy shift \textbf{e)} and intensity dependence \textbf{f)} versus threshold power for the $E_{k=0}$ of the lower polariton branch.}
\label{PD}
\end{figure}

In order to achieve condensation in the bottom of the dispersion we excite our sample with a 35$\mu$m spot at FWHM. Smaller excitation diameters result in condensation at a high $k$-vector due to a steeper potential profile induced by the repulsive exciton-exciton and exciton-polariton interactions \cite{wertz2010}. The optical excitation is provided by a 180fs pulsed Ti:Saph oscillator at a rep. rate of 80MHz. The non-resonant pulse excites an electron-hole plasma that rapidly relaxes to populate the polariton dispersion and uncoupled exciton reservoir. As the density of polaritons is raised the relaxation efficiency of polariton-polariton scattering is increased until the relaxation rate to the ground state of the lower polariton mode is sufficient to overcome the radiative decay of the system that results to a macroscopic ground state population \cite{kasprzak2006}. Fig\ref{PD}.a shows the polariton dispersion detected in the transmission geometry below threshold, where the upper polariton branch is also visible (the intensity of the upper polariton image is magnified 100 times). Gradually increasing the optical power we observe a threshold  ($P_{thr}=26\mu J/cm^{2}$) beyond which the emission shrinks in momentum space as shown with the $k$-space intensity profiles in figs\ref{PD}.a,b,c.  In figures\ref{PD}.b,c we have magnified the intensity above $|2| \mu m$ by 100 and 200 respectively, to clearly show that the excited population is strongly coupled following the lower polariton dispersion. The spectral properties of the $k=0$ energy state of the polariton dispersion indeed display the predicted features for polariton condensation, namely linewidth narrowing (fig.\ref{PD}d), a blueshift of the polariton mode (fig.\ref{PD}e), and a nonlinear increase in intensity (fig.\ref{PD}f). Beyond threshold, strong interactions between the polariton condensate and the co-located dark exciton reservoir further broadens the emission linewidth of the condensate as shown in fig.\ref{PD}d \cite{ring}. The continuous blueshift of the polariton mode above threshold is a good indication that the condensate has not entered the weak-coupling regime \cite{pillarbloch}. 
		
\begin{figure}
\includegraphics[bb=0 0 820 330,scale=0.52]{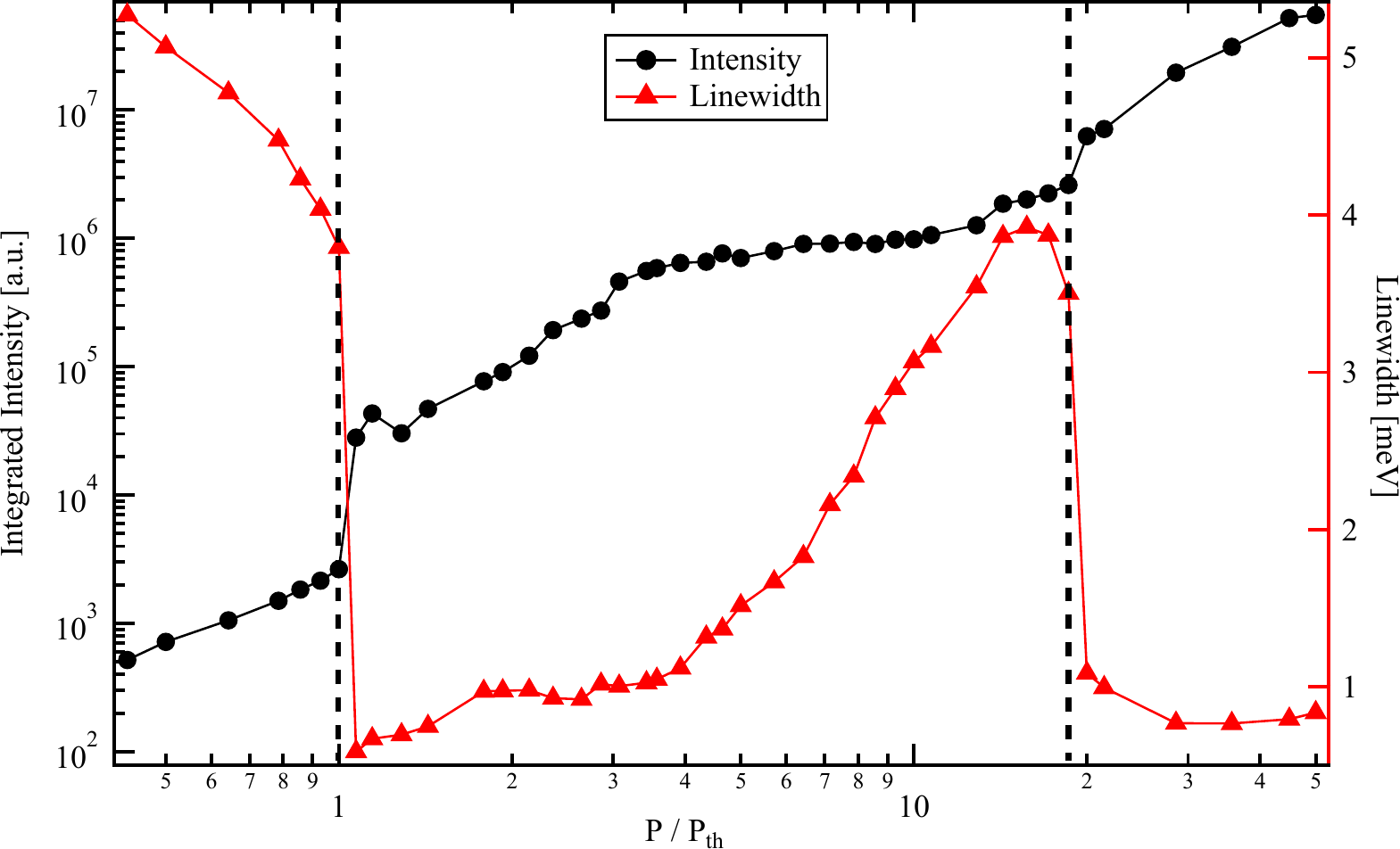}
\centering
\caption{ Integrated intensity and linewidth for increasing power marking the transition to the photon lasing regime.}
\label{Pulse}
\end{figure}

Moreover, unambiguous evidence of polariton condensation is provided by the observation of a second nonlinear threshold in the intensity upon which the system crosses to the weak coupling regime \cite{crossover}. To achieve the excitation density required to drive the system from the strong to weak coupling regime we focus our excitation beam down to a 2$\mu$m spot at FWHM. With increasing excitation density the non-linear polariton dispersion switches to a weakly coupled cavity mode. To record the evolution of the photoluminescence intensity between the two regimes we integrate the emission over the entire lower polariton branch. At approximately 20 times above the threshold for polariton condensation we observe a second nonlinear increase associated with a a further linewidth narrowing (fig.\ref{Pulse}). We note that integrating the emission over the lower polariton branch considerably broadens the measured emission linewidth and reduces the intensity offset between the linear and nonlinear plateau. This second nonlinearity is attributed to the transition to photon lasing, where the coherent emission originates from stimulated amplification due to population inversion as in the case of a VCSEL. We note here that this is the first observation of a second threshold in planar microcavity that is clearly associated with a substantial linewidth narrowing. 

In conclusion, we have presented experimental evidence of polariton condensation in a strain compensated GaAs-based cavity embedded with InGaAs QWs. This system exhibits all the important features of polariton condensation, namely a nonlinear increase of intensity, along with a linewidth narrowing and the observation of a second threshold nearly 20 times above the threshold, clearly marking this phase transition as polariton condensation in the strong coupling regime. As this type of strain compensated microcavity has been shown to suppress optical disorder and cross-hatched defects, it emerges as a favourable system for studying the exotic and ambivalent nature \cite{linear2014} of quantum fluid phenomena.

\begin{acknowledgments}
P.C., W. L. and P. L. acknowledge support by the EPSRC under Grant No. EP/309 F027958/1.
A.A. acknowledges funding from the POLATOM ESF research network.
The sample was grown by E.C. at the EPSRC National Centre for III-V Technologies, Sheffield, UK.
\end{acknowledgments}

%\bibliography{bibliography}

%

\end{document}